\documentclass[prb,cha,onecolumn,superscriptaddress,preprintnumbers,showpacs, amsmath,amssymb]{revtex4-1}
\usepackage{bm}
\usepackage[colorlinks=true,linkcolor=blue,citecolor=blue]{hyperref}
\usepackage{amsmath}
\usepackage{amssymb}
\usepackage{amsthm}
\usepackage{amsfonts}
\usepackage{enumerate}
\usepackage{latexsym}
\usepackage{ifpdf}
\usepackage{graphicx}
\usepackage{makeidx}
\expandafter\ifx\csname package@font\endcsname\relax\else
 \expandafter\expandafter
 \expandafter\usepackage
 \expandafter\expandafter
 \expandafter{\csname package@font\endcsname}
 \fi
\usepackage{color}
\usepackage{times}

\begin{document}

\title{Magnetism and electronic structure of  YTiO$_3$ thin films}
\author{Yanwei Cao}
\email{yc003@uark.edu}
\affiliation{Department of Physics, University of Arkansas, Fayetteville, Arkansas 72701, USA}

\author{P. Shafer}
\affiliation{Advanced Light Source, Lawrence Berkeley National Laboratory, Berkeley, California 94720, USA}

\author{Xiaoran Liu}
\affiliation{Department of Physics, University of Arkansas, Fayetteville, Arkansas 72701, USA}

\author{D. Meyers}
\affiliation{Department of Physics, University of Arkansas, Fayetteville, Arkansas 72701, USA}

\author{M. Kareev} 
\affiliation{Department of Physics, University of Arkansas, Fayetteville, Arkansas 72701, USA}

\author{S. Middey}
\affiliation{Department of Physics, University of Arkansas, Fayetteville, Arkansas 72701, USA}

\author{J. W. Freeland}
\affiliation{Advanced Photon Source, Argonne National Laboratory, Argonne, Illinois 60439, USA}

\author{E. Arenholz}
\affiliation{Advanced Light Source, Lawrence Berkeley National Laboratory, Berkeley, California 94720, USA}

\author{J. Chakhalian}
\affiliation{Department of Physics, University of Arkansas, Fayetteville, Arkansas 72701, USA}


\begin{abstract}

High-quality (001)-oriented (pseudo-cubic notation) ferromagnetic YTiO$_{3}$ thin films were epitaxially synthesized in a layer-by-layer way by pulsed laser deposition. Structural, magnetic and electronic properties were characterized by reflection-high-energy-electron-diffraction, X-ray diffraction, vibrating sample magnetometry, and element-resolved resonant soft X-ray absorption spectroscopy. To reveal ferromagnetism of the constituent titanium ions, X-ray magnetic circular dichroism spectroscopy was carried out using four detection modes probing complimentary  spatial  scale, which overcomes a  challenge of probing ferromagnetic titanium with pure Ti$^{3+}$(3$d^1$). Our work provides a pathway to distinguish between the  roles of titanium and A-site magnetic rare-earth cations in determining the magnetism in rare-earth titanates thin films and  heterostructures.   
\end{abstract}

\maketitle

\newpage

Recently the study of magnetism in titanates has attracted tremendous interest due to the variety of remarkable quantum many-body phenomena, i.e., spin ice states in magnetically frustrated pyrochlore titanates Ho$_2$Ti$_2$O$_7$ and Dy$_2$Ti$_2$O$_7$ with Ti 3$d^0$ electronic configuration, \cite{Science-2001-Steven} antiferromagnetism (AFM)-ferromagnetism (FM) crossover behavior of the ground states in the bulk perovskite rare-earth titanates $R$TiO$_3$ ($R$ = La ... Eu etc.) with Ti 3$d^1$ electronic configuration, \cite{RMP-1998-Imada,JPCM-2005-Zhou,NJP-2004-Moch,PRB-2007-Komarek,PRB-2010-Tak} strain-induced ferroelectric ferromagnet in EuTiO$_3$ films, \cite{nature-2010-Lee} and emerging interfacial ferromagnetism in the LaAlO$_3$/SrTiO$_3$ and GdTiO$_3$/SrTiO$_3$ heterostructures with a mixture of Ti 3$d^1$/3$d^0$ electronic configurations. \cite{NM-2012-Hwang,Nphy-2011-Millis,Nphy-2013-Gabay,PRB-2014-Ruh,MRS-2013-Coey,NCom-2010-GB,APL-2011-Moe,PRX-2012-Moe,PRB-2013-Jac} Particularly, the unexpected coexistence of ferromagnetism and superconductivity at the LaAlO$_3$/SrTiO$_3$ interface makes it intriguing  and challenging to understand the interfacial magnetism of titanates. \cite{Nphy-2013-Gabay,PRB-2014-Ruh,MRS-2013-Coey} On the other hand, in $R$TiO$_3$ most rare-earth elements have a partially filled 4$f$-shell with large total spin angular momentum (e.g., $S$ = 7/2 for Gd$^{3+}$), which are usually involved in magnetic interactions with the Ti$^{3+}$ sites. \cite{JPCM-2005-Zhou} The magnetism in titanates, in the bulk or at interfaces, can arise from either the rare-earth elements, oxygen vacancies, magnetic Ti$^{3+}$, or the combination of these. Therefore, element-resolved and valence-state specific investigations are fundamentally important and imperative not only for greater understanding the role of Ti$^{3+}$ in the emergence of magnetism but also for  applications to spintronics based on titanate heterostructures.

Experimentally, however, there are several obstacles on the way to investigating magnetic order of Ti$^{3+}$. First, only very few techniques are capable of probing the element-resolved magnetism on  nanoscale. To this end, circularly polarized resonant soft X-ray absorption spectroscopy, used in this work, is  one of the most powerful available probes, with the total electron yield (TEY, surface or interface sensitive) mode having been proven to be  suitable  to investigate the interfacial magnetism. \cite{RMP-2014-JC,NP-2006-JC,Science-2007-JC,NCom-2010-GB,Nmat-2013-Lee} At the same time, surprisingly, it was found that in the LaMnO$_3$/SrTiO$_3$ and LaAlO$_3$/SrTiO$_3$ heterostructures, the circular dichroism signal of interfacial Ti$^{3+}$/Ti$^{4+}$ with TEY detection mode at Ti $L_{2,3}$-edge is very weak ($\sim$ 1\% of absorption spectra signal). \cite{NCom-2010-GB,Nmat-2013-Lee} Additionally, synthesis of high-quality samples with pure Ti$^{3+}$ (3$d^1$) is challenging due to the rapid conversion of magnetic Ti$^{3+}$ to non-magnetic Ti$^{4+}$ in low vacuum. \cite{APL-2002-Ohto,APL-2006-Chae,APL-2013-Misha} To fully understand the intriguing quantum many-body phenomena in magnetic Ti$^{3+}$ systems, synthesis and experimental investigation of the magnetism of pure Ti$^{3+}$  is essential and yet thus far lacking. \cite{NCom-2010-GB,Nmat-2013-Lee}

In this Letter, we study ferromagnetic Mott insulator YTiO$_3$ films (YTO) as a prototypical rare-earth titanate system. The structural and electronic quality of YTO films were confirmed by reflection-high-energy-electron-diffraction (RHEED), X-ray diffraction (XRD), vibrating sample magnetometer (VSM) in Physical Properties Measurement System (PPMS), and resonant soft X-ray absorption spectroscopy (XAS). We thoroughly investigated the circular magnetic dichroism (XMCD)\ of titanium ferromagnetism in luminescence yield, reflection, fluorescence yield, and electron yield detection modes. Our work directly demonstrated the presence of long range ferromagnetic order of the constituent titanium ions.

\begin{figure*}[htp!]\vspace{-0pt}
\includegraphics[width=0.6\textwidth]{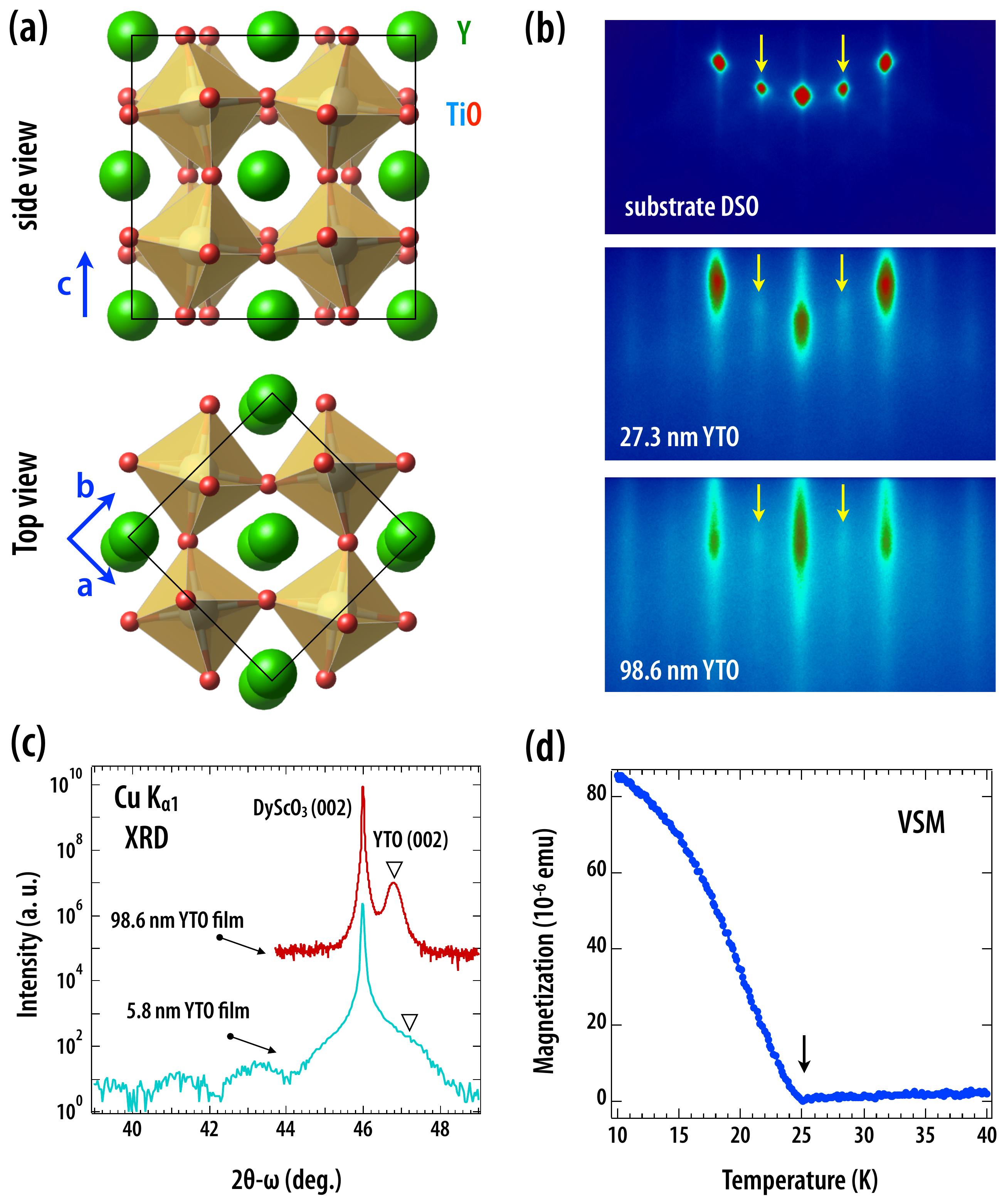}
\caption{\label{} (a) Schematic views of bulk YTiO$_{3}$ unit cell (orthorhombic notation). (b) RHEED patterns of YTO film ($\sim$ 98.6 nm thickness) during growth. Yellow arrows indicate the half-order-peaks. (c) XRD data of YTO films on DyScO$_3$ substrates. The main substrate peak corresponds to 3.946 \AA ~(pseudo-cubic notation). Due to the smaller lattice parameter of bulk YTO than that of substrate, the superlattices are under tensile strain. The black triangles suggest the peaks of films under almost full ($\sim$ 5.8 nm) and relaxed ($\sim$ 98.6 nm) epitaxial strain.  (d) Temperature-dependent magnetization of YTO film with a subtraction of constant background. The magnetization was measured during zero field warming up process after field cooling (2000 Oe) by VSM in PPMS. The black arrow indicates the critical temperature of ferromagnetic transition near 25 K.}
\end{figure*}

High-quality (001)-oriented (pseudo-cubic notation) YTO films were grown in a layer-by-layer fashion on DyScO$_3$ (001)-oriented (pseudo-cubic notation, corresponding to (110)-orientation in orthorhombic notation) single crystal substrates by pulsed laser deposition (PLD) under the same growth conditions reported elsewhere. \cite{APL-2013-Misha} For this work, YTO was carefully selected  as a prototype for three reasons: (1) in contrast to other rare-earth elements, the total spin angular momentum of Y$^{3+}$ with 4$d^0$ electronic configuration is almost zero thus ruling out the contribution of the magnetic rare-earth element, (2) the ferromagnetic transition temperature  of bulk YTO is comparatively high (27 - 30~K), \cite{APL-2006-Chae,PRB-2009-Knafo,PB-2003-Suzuki} and (3) bulk YTO has a large Mott gap ($\sim$1.2~eV) compared for example to easily oxidized and metallic LaTiO$_3$ ($\sim$ 0.2~eV).\cite{PRB-1995-Oki} 

Due to the small ionic radius of Y$^{3+}$, bulk YTO has a highly distorted GdFeO$_3$-type orthorhombic structure ($Pbnm$ space group) [see Fig.~1(a)] with lattice constants of $a$~=~5.679\AA, $b$~=~5.316\AA, and $c$~=~7.611\AA, \cite{PRB-2002-Nakao,JPCM-2007-Loa} and shows intriguing orbital ordering \cite{PRB-1998-Sawada,PRB-2002-Nakao,EPL-2005-Okatov} with magnetic and orbital excitations. \cite{NJP-2004-Tanaka,PRL-2006-Ulrich,PRB-2008-Ulrich}  In this work, to monitor the quality of YTO films, during
growth RHEED patterns were recorded, as illustrated in Fig.~1(b). As seen, the sharp RHEED patterns with expected orthorhombic half-order reflections [yellow arrows in Fig.~1(b)] indicate a two-dimensional growth mode of YTO films. Structural quality of the thick ($\sim$ 98.6 nm) and ultra-thin ($\sim$ 5.8 nm) films were further confirmed by XRD using Cu K$_{\alpha1}$ radiation shown in Fig.~1~(c). To characterize the magnetic properties of thick YTO films, the magnetization versus temperature curve was measured with VSM in a PPMS. As shown in Fig.~1(d), as anticipated there is a clear ferromagnetic transition around 25~K for YTO film, showing excellent agreement with the bulk value of 27-30~K in single crystals \cite{PRB-2009-Knafo,PRB-2002-Nakao,PB-2003-Suzuki}.

\begin{figure}[t]\vspace{-0pt}
\includegraphics[width=0.4\textwidth]{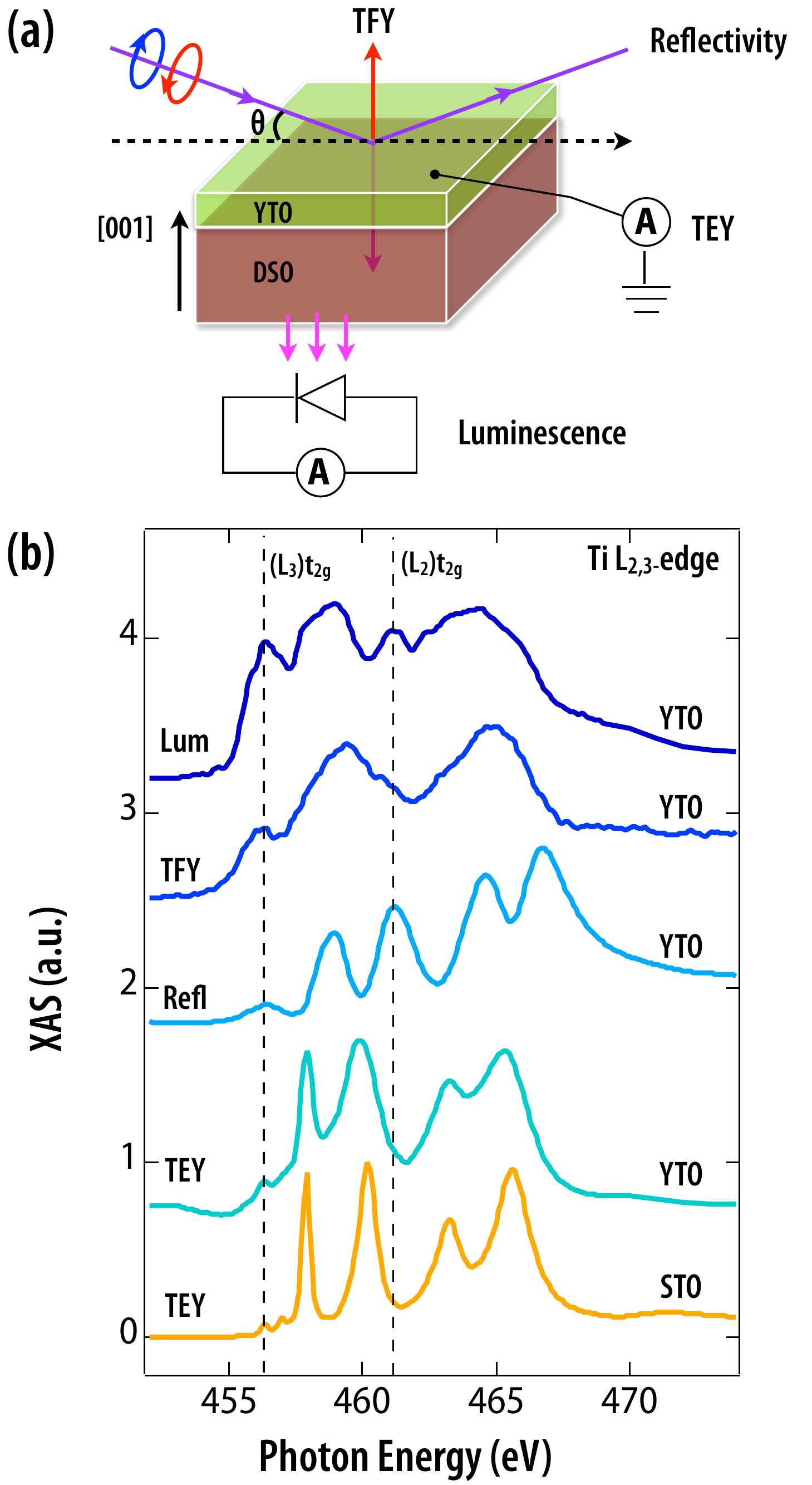}
\caption{\label{} (a) Schematic of XAS/XMCD signal detection using four different modes. During measurements, the grazing angles $\theta$ were kept at 20$^{\circ}$ and 10$^{\circ}$ for the beamlines 4.0.2 (ALS) and 4-ID-C (APS), respectively. Here, the arrow [001] (pseudo-cubic notation) indicates the growth direction of YTO films. (b) Ti $L_{2,3}$-edge XAS on YTO film using four modes-luminescence yield (Lum, 11K), total fluorescence yield (TFY, 15 K), reflectivity (Refl, 15K), and total yield electron (TEY, temperature 11K). The reference sample SrTiO$_3$ single crystal (5 $\times$ 5 $\times$ 0.5 mm$^3$) was measured at room temperature. XAS spectra of perovskite titanates is almost independent of temperature. \cite{PRL-2005-Hav,JPSJ-2007-Ari} The dashed lines indicate the energy positions of t$_{2g}$ state of Ti$^{3+}$ 3$d$ band. The two small pre-edge peaks of STO spectra (bottom, orange curve) at $\sim$457 eV were attributed to the multiplet core hole-$d$ electron interactions. \cite{PCM-2002-Hen}}
\end{figure}

\begin{figure}[t]\vspace{-0pt}
\includegraphics[width=0.45\textwidth]{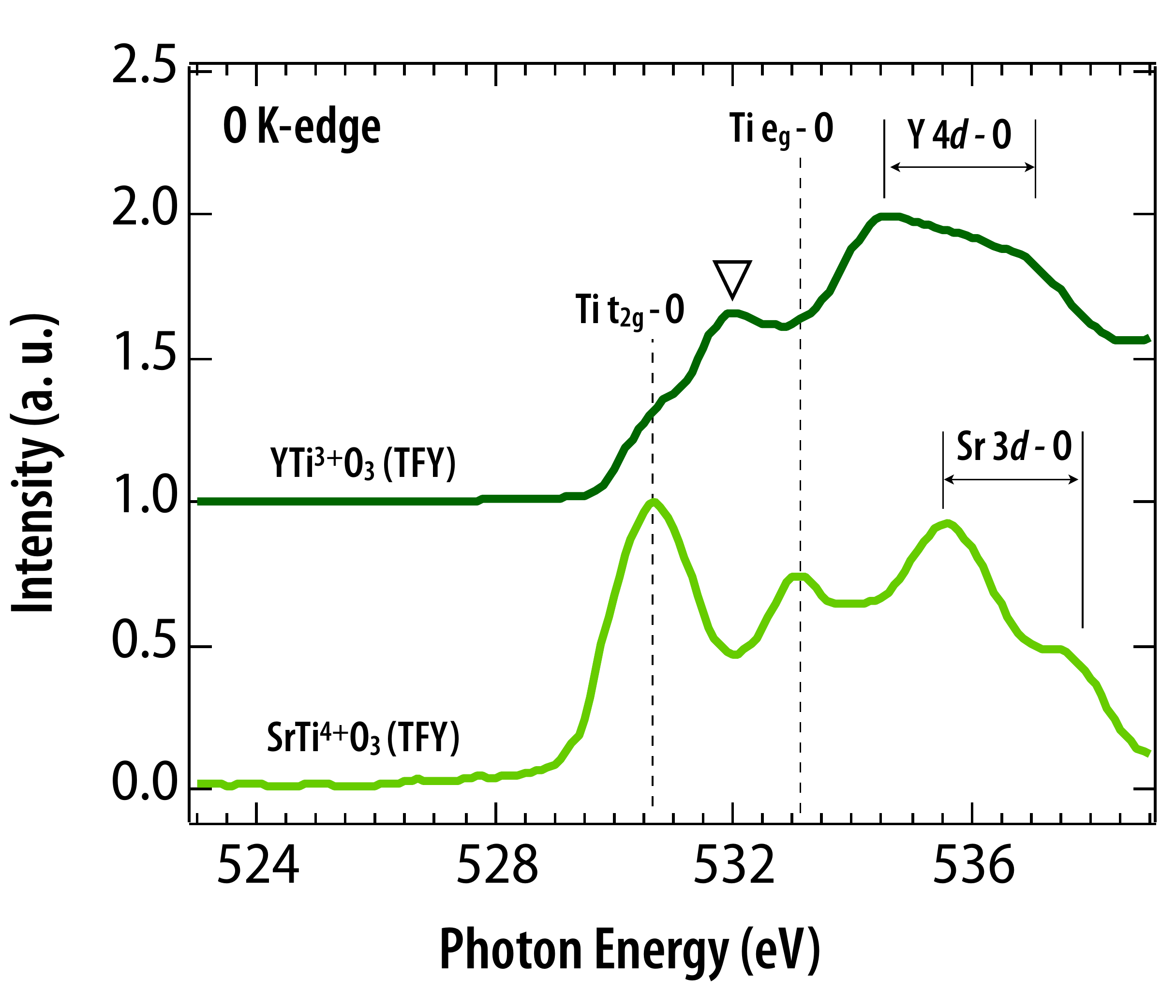}
\caption{\label{} O $K$-edge XAS on YTO film at 15 K. The triangle suggests the second peak of O $K$-edge spectra on YTO film. The spectra of SrTiO$_3$ single crystal as a reference was measured at room temperature. }
\end{figure}

\begin{figure}[t]\vspace{-0pt}
\includegraphics[width=0.45\textwidth]{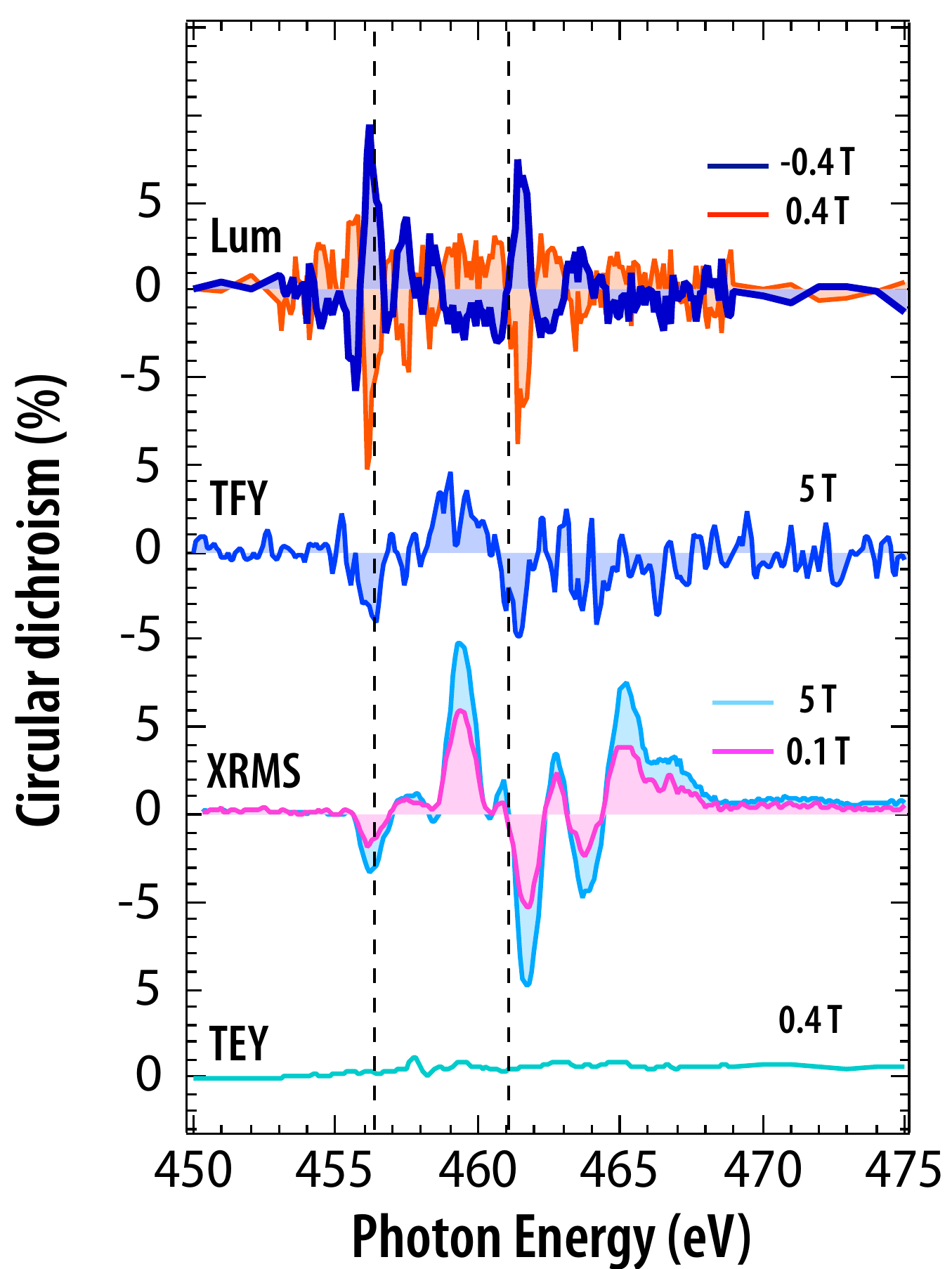}
\caption{\label{} Circular dichroism of YTO film under four detection modes - luminescence (Lum, 11K), total fluorescence yield (TFY, 15 K), reflectivity (XRMS, 15K), and total yield electron (TEY, temperature 11K). The dashed lines near 456 and 461 eV indicate the features derived from the t$_{2g}$ state of Ti$^{3+}$ 3$d$ band. The degree of circular polarization is near 100\%. The direction of the applied magnetic field is parallel to the sample surface. Comparing with the magnetization versus magnetic field of bulk YTO,\cite{JPCM-2005-Zhou} the magnetic field-dependent XRMS intensity indicates the magnetization of the YTO film under 0.4 T magnetic field is near saturation.} 
\end{figure}

To further investigate the electronic and magnetic structures of YTO film, XAS measurements using four detection modes were carried out at two synchrotron facilities, beamline 4.0.2 (using the vector magnet) of the Advanced Light Source (ALS, Lawrence Berkeley National Laboratory) and beamline 4-ID-C of the Advanced Photon Source (APS, Argonne National Laboratory). As schematically illustrated in Fig.~2(a), these four detection modes are total electron yield (TEY, surface sensitive with 2-10 nm probing depth, measured at ALS with a 20$^{\circ}$ incident angle), \cite{PRB-2000-Gota,PRB-1997-Pompa} total fluorescence yield (TFY, bulk probing depth $\sim$ 20 nm, measured at APS with a 10$^{\circ}$ incident angle), \cite{PRB-1997-Pompa} reflectivity (surface sensitive, measured at APS with a 10$^{\circ}$ incident angle), \cite{JPCM-2007-Freeland} and new luminescence yield mode (Lum, bulk probing depth $\sim$ 50 nm, measured at ALS with a 20$^{\circ}$ incident angle). \cite{PRL-1995-Chen,JESRP-2004-Huang,PRB-2012-Meinert,SP-2012-Kach,JPD-2007-Kal} Since pure Ti$^{3+}$ is challenging to stabilize and it  easily  converts to  non-magnetic Ti$^{4+}$  on the YTO film surface, only bulk-sensitive TFY and luminescence detection modes are able to probe the properties of the Ti$^{3+}$.
In the past , however, TFY signal has been known to distort the  line-shape due to the saturation and self-absorption effects, \cite{PRB-1993-Eis} generally produces a very small absorption signal and even smaller magnetic dichroism at X-ray energies around Ti $L_{2,3}$-edge ($\sim$460 eV) due to the low number of fluorescence transitions. On the other hand, in a \textit{luminescence yield} measurement \cite{PRL-1995-Chen,JESRP-2004-Huang,PRB-2012-Meinert,SP-2012-Kach,JPD-2007-Kal} the X-ray beam transmitted through the YTO thin film is converted to visible light in the substrate. This luminescence signal can be detected with a diode behind the sample as a function of X-ray photon energy and is free of saturation and self-absorption effects. As the result, it is a direct and primarily bulk sensitive probe.

With the confirmed structural quality, we investigated the electronic structures of YTO films  by XAS at the Ti $L_{2,3}$-edge. As seen in Fig.~2(b), in contrast with the reference Ti$^{4+}$ spectra of bulk SrTiO$_3$ (STO) with four characteristic peaks,\cite{Nmat-2013-Lee} the spectra of Ti$^{3+}$ in YTO film shows a distinct lineshape (i.e. two main peaks) with a $\sim$ 2 eV chemical shift to lower energy \cite{PRL-2005-Hav} -  the difference  is mainly connected to the crystal field splitting and the Ti 3$d^{1}$ electronic configuration of YTO. In addition, due to the oxidation of Ti$^{3+}$ to Ti$^{4+}$ on the YTO film surface causing a mixture of Ti$^{3+}$/Ti$^{4+}$ after being taken out of the vacuum chamber and exposed to the atmosphere during the delivery process, a significant Ti$^{4+}$ contribution is observed by \textit{surface-sensitive} TEY and reflectivity modes. For the interior unit cells, the recently developed bulk  sensitive luminescence mode \cite{PRB-2012-Meinert,SP-2012-Kach} shows much clearer signal than that of the TFY mode characteristic of  pure Ti$^{3+}$.

To further investigate the difference in electronic structure caused by covalency between Ti$^{3+}$-O and Ti$^{4+}$-O bonds, O $K$-edge spectra were measured. To  get a\ fingerprint of Ti$^{4+}$ we acquired O $K$-edge spectra on a STO\ single crystal.  As seen in Fig.~3, due to the  hybridization between Ti and O ions, the near-edge two main peaks ($\sim$ 530.5 and 533 eV, marked by the dashed lines)   derived from Ti t$_{2g}$ and Ti e$_{g}$ bands are present. \cite{JPSJ-2007-Ari,PRB-1989-Groot} In contrast to  STO, the O $K$-edge spectra of the YTO film shows that the first peak ($\sim$ 530.5 eV) is strongly suppressed whereas the second peak ($\sim$ 532 eV, marked by the triangle in Fig.~3) is  shifted to lower energy and is enhanced; this  result  agrees well with the previously reported spectra of bulk YTO. \cite{JPSJ-2007-Ari} To identify these features, we recap  that  compared to   the Ti t$_{2g}$ and e$_{g}$ band splitting  on STO\  driven by the large cubic crystal field, the recent YTO\ band structure calculation predicts the presence of an occupied lower Hubbard band (LHB) and empty upper Hubbard band (UHB) with partial  mixing  of  Ti t$_{2g}$ and e$_{g}$ bands.\cite{PRB-2014-Him} Experimentally, previous  resonant X-ray \textit{inverse} photoemission data attributed the second peak ($\sim$ 532 eV) seen in our O $K$-edge spectra to the UHB of YTO. \cite{JPSJ-2007-Ari} On the other hand, for the two peaks with higher energy ($\sim$ 536 eV), they are mainly derived from the hybridizations between the $d$ bands of A-site ions (Y and Sr) and the O 2$p$ bands (see Fig.~3). \cite{JPSJ-2007-Ari}

To probe magnetism,  we carried out XMCD measurements at the Ti \textit{L}$_{2,3}$-edge [see Fig.~2(a)] to reveal the magnetic properties of the YTO film, as shown in Fig.~4. Left and right circularly polarized soft X-rays were tuned to the Ti $L_{2,3}$-edge and recorded in TEY, \cite{PRB-2000-Gota,PRB-1997-Pompa} TFY, \cite{PRB-1997-Pompa} reflectivity (X-ray resonant magnetic scattering, XRMS) \cite{JPCM-2007-Freeland} and Luminescence \cite{PRL-1995-Chen,JESRP-2004-Huang,PRB-2012-Meinert,SP-2012-Kach,JPD-2007-Kal} detection modes below 15 K and in an applied magnetic  field. First, we  point out that despite the surface sensitivity of TEY mode, the non-magnetic T$^{4+}$ (3$d^{0}$)\ ions present on the surface of YTO film do not  contribute  to the magnetism in  YTO film and as such  the  XMCD signal is  naturally absent in the TEY mode. For another surface-sensitive mode, XRMS, the XMCD signal near the Ti absorption edge represents the convoluted contribution from both chemical and magnetic scattering, and requires extensive modeling to identify the relevant Ti$^{3+/4+}$ states involved. \cite{JPCM-2007-Freeland,PRL-1990-Kao}  For the bulk-sensitive TFY mode, on the other hand, a small XMCD signal is visible but its spectral features are difficult to differentiate due to the low fluorescence yield intensity in this energy range.  Finally, by using the  luminescence detection method we were able to  obtain a strong  ferromagnetic signal on Ti.
As seen in Fig. 4, the circular dichroism signal is  strong ($\sim$10\%), and its sign (negative or positive)  can be flipped by reversing the orientation of the external magnetic field. To  emphasize, the circular dichroism  detected in the bulk sensitive luminescence mode  under  small magnetic field of $\pm 0.4$ T lends strong support to the ferromagnetic Ti$^{3+}$ state, whose spectral features are consistent with the suggested ferromagnetic titanium in LaMnO$_3$/SrTiO$_3$ and LaAlO$_3$/SrTiO$_3$ interfaces  and the theoretically predicted  Ti 3$d^{1}$ electronic configuration. \cite{NCom-2010-GB,Nmat-2013-Lee,PRB-1991-Laan}  However, due to the insufficiently separated Ti $L_{2}$- and $L_{3}$-edge peaks, the pronounced shoulders, and the considerable contribution of the magnetic dipole term for solids lacking cubic symmetry, \cite{JAP-1994-OB,PRB-1994-OB} the application of the XMCD sum rules for magnetic $R$TiO$_3$ is very difficult.

In summary, we developed layer-by-layer growth of high-quality Mott insulator YTiO$_3$ films and investigated  their electronic  structure and  ferromagnetism of the Ti and O ions. A combination of RHEED, XRD, VSM and XAS confirmed the proper structural, chemical, magnetic and electronic quality. The ferromagnetism of the constituent Ti ions was directly observed with X-ray circular dichroism measurements in the luminescence detection mode. Our work provides a pathway to disentangle the independent roles of Ti and the A-site  magnetic rare-earth ions in the perovskite titanates, and is  important for studies of interfacial magnetism  in $R$TiO$_{3}$-based heterostructures.

The authors deeply acknowledge the discussions with Se Young Park. Research at  the University of Arkansas is funded in part by the Gordon and Betty Moore Foundation EPiQS Initiative through Grant GBMF4534 and synchrotron  work was supported by the DOD-ARO under Grant No. 0402-17291. The Advanced Light Source is supported by the Director, Office of Science, Office of Basic Energy Sciences, of the U.S. Department of Energy under Contract No. DE-AC02-05CH11231. This research used resources of the Advanced Photon Source, a U.S. Department of Energy (DOE) Office of Science User Facility operated for the DOE Office of Science by Argonne National Laboratory under Contract No. DE-AC02-06CH11357.

\end{document}